\documentclass[5p,sort&compress]{elsarticle}

\usepackage{lineno,hyperref}
\modulolinenumbers[5]

\usepackage{amsmath}
\usepackage{amssymb}

\usepackage{hyperref}

\journal{ }

\begin{document}

\begin{frontmatter}

\title{On the nature of the Newton's gravitational constant and the possible quantum-field theory of gravitation}

\author{E.A.~Pashitskii}
\ead{pashitsk@iop.kiev.ua}

\author{V.I.~Pentegov \corref{corr}}
\ead{pentegov@iop.kiev.ua}
\cortext[corr]{Corresponding author}

\address{Institute of Physics, NAS of Ukraine, 46 Nauki Avenue, Kiev, 03028, Ukraine}

%
%

\begin{abstract}
On the basis of the coincidence of the physical dimensions (in natural units $\hbar = c = 1$) of the Newton's gravitational constant $G_{N} $ and the phenomenological Fermi constant $G_{F} $ for weak interaction, it is suggested that there is a certain similarity between weak forces, which are caused by the exchange of massive intermediate vector bosons with spin $S=1$, and ``superweak'' gravitational forces that can be caused by the exchange of ``supermassive'' hypothetical tensor bosons with spin $S=2$. By analogy with how the masses of intermediate bosons in the theory of electroweak interaction arise as a result of spontaneous breaking of the gauge symmetry of the electromagnetic field due to its interaction with the nonlinear scalar Higgs field, the masses of hypothetical tensor bosons carrying gravitational interaction can also arise as a result of spontaneous breaking of gauge symmetry of the massless gravitons when they interact with a fundamental nonlinear scalar field in a flat 4-dimensional space-time.
\end{abstract}


\end{frontmatter}


\section{Introduction}
One of the main obstacles to the construction of the renormalizable quantum field theory of gravity is that the corresponding interaction constant, the Newton's gravitational constant $G_{N} $, is a dimensional one $\left[ {{G_N}} \right] = \left[ {{e^2}/{m^2}} \right]$. It should be noted that in natural units $\hbar = c = 1$ (where $\hbar $ is the Planck constant and $c$ is the speed of light) the dimension of $G_{N} $ coincides with the dimension of the phenomenological Fermi constant $G_{F} $ for weak interaction, for which the renormalizability problem was successfully solved in the Weinberg-Salam \cite{Weinberg1996} theory of the unified electroweak interaction. Within the framework of this theory, it was shown that weak interaction is caused by the exchange of virtual massive intermediate bosons $W^{\pm } $ and $Z^{0} $ whose masses arise due to the spontaneous breaking of the gauge symmetry of the vector massless electromagnetic field with spin $S=1$ as a result of its interaction with the nonlinear scalar Higgs field \cite{Higgs1964a}.

In this connection, based on the coincidence of the physical dimensions of the constants $G_{N} $ and $G_{F} $, it was suggested in \cite{Pashitskii2017} that the ``superweak'' gravitational interaction is due to the exchange of virtual ``supermassive'' tensor bosons with spin $S=2$, but the question of the origin of such bosons with a mass ${M_T}$ on the order of the Planck mass ${M_P} = \sqrt {{{\hbar c} \mathord{\left/ {\vphantom {{\hbar c} {{G_N}}}} \right. \kern-\nulldelimiterspace} {{G_N}}}} $ was not discussed in \cite{Pashitskii2017}. We should note that earlier on the basis of a comparison of the constants $G_{N} $ and $G_{F} $ the possibility of ``scaling'' between the gravitational and weak interactions was considered in \cite{Onofrio2013,Onofrio2014}.

In this paper it is shown that the generation of a finite mass of tensor bosons can be caused by the spontaneous breaking of the gauge symmetry of the tensor field of massless gravitons -- the quanta of gravitational field. Gravitons were predicted by Einstein \cite{Einstein1916} on the basis of the general theory of relativity (GTR) as gravitational waves, and were discovered recently in the LIGO (Laser Interferometer Gravitational-Wave Observatory)  experiments \cite{Abbott2016}. However, such a mechanism of mass generation due to the interaction of gravitons with a nonlinear scalar field having a nonzero vacuum expectation value is possible only for a weak gravitational field, which in the linear approximation with respect to small perturbations of the metric tensor has local gauge symmetry in a flat 4-dimensional space-time \cite{Weinberg1972}.

At the same time, according to the general principles of GTR, the presence of a nonlinear scalar field with a finite potential energy density, which determines the vacuum energy density $\lambda _{0} $ and the cosmological constant $\Lambda _{0} =\kappa \lambda _{0} $ (where $\kappa  = 8{\pi ^2}{{{G_N}} \mathord{\left/ {\vphantom {{{G_N}} {{c^4}}}} \right. \kern-\nulldelimiterspace} {{c^4}}}$ is the Einstein's gravitational constant) in the equations of general relativity, must create a negative scalar curvature $R_{0} =-4\Lambda _{0} $ of the 4D space-time. In such a curved space, the gauge symmetry of the gravitational field is initially violated, and the tensor boson mass generation mechanism analogous to the Higgs mechanism \cite{Higgs1964a} for intermediate vector bosons \cite{Weinberg1996} is impossible.

In connection with this, along with string theory in multidimensional anti-de Sitter spaces \cite{Maldacena1998,Aharony2000,Aharony2008}, for a long time, beginning with \cite{DeAlfaro1980}, attempts were made to construct field-theoretical models of quantum gravity in curvilinear 4D space-time by combining nonlinear Yang-Mills equations \cite{Yang1954} in quantum chromodynamics (QCD) with nonlinear equations of general relativity \cite{Westman2014} or with its modified version, which is characterized by a quadratic dependence of the Lagrangian on the scalar curvature $ \propto {R^2}$ \cite{Holdom2016}.

In this paper, we propose a different approach based on the effective exclusion of gravity and curvature of 4D space-time within the framework of a possible version of the quantum field theory of exchange gravitational interaction in which the intensity of this interaction is determined by the double convolution of the retarded Green function of massive tensor bosons with metric tensors. Such a convolution may be written as ${\tilde G_N}(p) = {G_N} \cdot \left( {1 - {{{p^2}} \mathord{\left/ {\vphantom {{{p^2}} {M_T^2}}} \right. \kern-\nulldelimiterspace} {M_T^2}}} \right)$, where ${p^2} = \left( {{\omega ^2} - {{\vec k}^2}} \right)$ is the 4D energy-momentum vector squared, and reduces to Newton's gravitational constant $G_{N} $  only in the static long-wavelength limit, when $p^{2} \to 0$, or for massless gauge fields with the energy of quanta  $\omega \left( {\vec k} \right) = \left| {\vec k} \right|$, where $\vec{k}$ is the 3D momentum (here and below we assume $\hbar = c = 1$).

However, for a vacuum filled with a nonlinear massive scalar field with a mass of bosons $M_{B} $, the spectrum of quantum fluctuations at absolute zero of temperature has a $\delta $-shaped peak at a frequency corresponding to the energy of quanta ${\Omega _B}\left( {\vec k} \right) = \sqrt {M_B^2 + {k^2}} $. Under the condition that the mass of scalar bosons $M_{B} $ is equal to the mass of tensor bosons ${M_T}$, the double convolution of the Green function of the tensor bosons is zero, so that in this case gravity is turned off. As a result, the 4D space-time averaged over the quantum fluctuations of the vacuum with frequency ${\Omega _B}\left( {\vec k} \right)$ on scales exceeding the Planck length and time scales ${l_P} = {t_P} = {1 \mathord{\left/ {\vphantom {1 {{M_P}}}} \right. \kern-\nulldelimiterspace} {{M_P}}}$ is effectively flat. In this case, the gauge symmetry of the weak gravitational field is restored, which leads to the possibility of a mass generation mechanism for tensor bosons due to the spontaneous breaking of the gauge symmetry of massless gravitons when they interact with the nonlinear scalar field having a nonzero vacuum average.

\section{\label{Sec2} Spontaneous breaking of the gauge symmetry of gravitons and the mechanism of mass generation of tensor bosons in a flat 4D space-time }

We consider the problem of spontaneous breaking of the local gauge symmetry of massless gravitons with spin $S=2$ due to their interaction with a nonlinear scalar field $\varphi $ whose Lagrangian has the following form:
\begin{equation} \label{1} {L_\varphi } = \frac{{{g^{\mu \nu }}}}{2}{\partial _\mu }\varphi  \cdot {\partial _\nu }\varphi  + \frac{{{\mu ^2}}}{2}{\varphi ^2} - \frac{{{g^2}}}{4}{\varphi ^4},  \end{equation}
where $g^{\mu \nu } $ is the metric tensor of 4D space-time, $\mu =im$ is the ``imaginary mass'' parameter, and $g$ is the non-linearity parameter of the scalar field. In the ground state, the scalar field is characterized by a nonzero vacuum expectation value of the amplitude ${\varphi _0} = {\mu  \mathord{\left/ {\vphantom {\mu  g}} \right. \kern-\nulldelimiterspace} g}$ and the density of the potential energy ${U_0}{{ = {\mu ^2}\varphi _0^2} \mathord{\left/ {\vphantom {{ = {\mu ^2}\varphi _0^2} 4}} \right. \kern-\nulldelimiterspace} 4}$, which determines the energy density of the physical vacuum $\lambda _{0} =U_{0} $, while the excitations of such a field are scalar bosons with mass ${M_B} = \mu \sqrt 2 $.

If we assume that for such a fundamental scalar field the mass of bosons is equal to the Planck mass, ${M_B} = {M_P} = {1 \mathord{\left/ {\vphantom {1 {\sqrt {{G_N}} }}} \right. \kern-\nulldelimiterspace} {\sqrt {{G_N}} }}$, and the energy density is equal to the limiting Planck value ${U_0} = {{{\mu ^4}} \mathord{\left/ {\vphantom {{{\mu ^4}} {4{g^2}}}} \right. \kern-\nulldelimiterspace} {4{g^2}}} = M_P^4$, then for the nonlinearity constant and the vacuum average of the scalar field we obtain $g = {1 \mathord{\left/ {\vphantom {1 4}} \right. \kern-\nulldelimiterspace} 4}$ and $\varphi _{0} =2M_{P } \sqrt{2} $.

According to GTR, a weak gravitational field in the linear approximation with respect to small perturbations of the metric tensor is described by the following tensor equation (see \cite{Weinberg1972}):
\begin{equation} \label{ZEqnNum927569} \square ^{2} h_{\mu \nu } -\frac{\partial ^{2} }{\partial x^{\lambda } \partial x^{\mu } } h_{\nu }^{\lambda } -\frac{\partial ^{2} }{\partial x^{\lambda } \partial x^{\nu } } h_{\mu }^{\lambda } +\frac{\partial ^{2} }{\partial x^{\mu } \partial x^{\nu } } h_{\lambda }^{\lambda } =0.  \end{equation}

This equation describes the free tensor field of massless gravitons in the absence of gravitating masses.

For the infinitesimal transformation of the 4D space-time coordinates ${x^\mu } \to {x^\mu } + {\varepsilon ^\mu }\left( x \right)$, where the smallness of the displacement ${\varepsilon ^\mu }\left( x \right)$ is determined by the condition ${{\partial {\varepsilon ^\mu }} \mathord{\left/ {\vphantom {{\partial {\varepsilon ^\mu }} {\partial {x^\nu }}}} \right. \kern-\nulldelimiterspace} {\partial {x^\nu }}} \sim \left| {{h_{\mu \nu }}} \right| \ll 1$, the metric tensor and the amplitude of the field \eqref{ZEqnNum927569} take the form:
\begin{equation} \label{3} g'^{\mu \nu } =\frac{\partial x'^{\mu } }{\partial x^{\lambda } } \frac{\partial x'^{\nu } }{\partial x^{\rho } } g^{\lambda \rho } ; \; h'^{\mu \nu } =h^{\mu \nu } -\frac{\partial \varepsilon ^{\mu } }{\partial x^{\lambda } } g^{\lambda \nu } -\frac{\partial \varepsilon ^{\nu } }{\partial x^{\rho } } g^{\rho \mu } ,  \end{equation}

For a flat 4D space-time with a Minkowski metric tensor, by virtue of the relation $\varepsilon _{\mu } =\varepsilon ^{\nu } g_{\mu \nu } $, we have:
\begin{equation} \label{ZEqnNum269772} {h'_{\mu \nu }} = {h_{\mu \nu }} - {{\partial {\varepsilon _\mu }} \mathord{\left/ {\vphantom {{\partial {\varepsilon _\mu }} {\partial {x^\nu }}}} \right. \kern-\nulldelimiterspace} {\partial {x^\nu }}} - {{\partial {\varepsilon _\nu }} \mathord{\left/ {\vphantom {{\partial {\varepsilon _\nu }} {\partial {x^\mu }}}} \right. \kern-\nulldelimiterspace} {\partial {x^\mu }}}.  \end{equation}

By substituting \eqref{ZEqnNum269772} into \eqref{ZEqnNum927569}, it is easy to see that $h'_{\mu \nu } $, just like $h_{\mu \nu } $, is a solution of equation \eqref{ZEqnNum927569}, which is a consequence of the exact local gauge symmetry of the weak gravitational field in flat 4D space-time.

By analogy with the Lagrangian of the interaction of scalar fields with ``color'' vector fields $A_\mu ^\alpha \left( x \right)$ given in \cite{Weinberg1996}, we consider the corresponding Lagrangian of the inter\-action of a scalar field $\varphi $ with a tensor gauge field ${h_{\mu \nu }}\left( x \right)$:
\begin{equation} \label{ZEqnNum805752} {L_\varphi } =  - \frac{1}{2}{\left( {{\partial _\mu }\varphi  - \tilde g \cdot \varphi  \cdot {h_{\mu \nu }}} \right)^2} ,  \end{equation}
where $\tilde{g}$ is the interaction constant of tensor and scalar fields.

Assuming that the gauge symmetry of the tensor fields is violated by the nonzero vacuum expectation value $\varphi _{0} $ of the scalar field $\varphi (x) = {\varphi _0} + \phi \left( x \right)$, with $\left| \phi  \right| \ll {\varphi _0}$, and neglecting the small cross terms of the type $\left( {\phi  \cdot {h_{\mu \nu }}} \right)$, we reduce \eqref{ZEqnNum805752} to the following form:
\begin{equation} \label{6} {L_\varphi } =  - \frac{1}{2}{\partial _\mu }\phi  \cdot {\partial ^\mu }\phi  - \frac{1}{2}{\tilde g^2}\varphi _0^2 \cdot {h_{\mu \nu }}{h^{\mu \nu }},  \end{equation}
where the quantity $M_{T} =\tilde{g}\cdot \varphi _{0} $ plays the role of a mass of tensor bosons that arises as a result of spontaneous breaking of the gauge symmetry of massless gravitons due to their interaction with the nonlinear scalar field $\varphi $ having a nonzero vacuum average.

\section{Similarity between weak and gravitational forces and the exchange nature of the gravitational interaction}

In the theory of electroweak interaction \cite{Weinberg1996}, the charged $W^{\pm } $ and neutral $Z^{0} $ intermediate bosons, in spite of the difference in their masses, are the three components of a single massive vector field with spin $S=1$ that arises from spontaneous breaking of the gauge symmetry of the electromagnetic field due to its interaction with the nonzero vacuum average of the nonlinear scalar Higgs field \cite{Higgs1964a}.

The Fourier representation of the imaginary-time Gre\-en function of a massive vector field has the form (see \cite{Berestetskii1982}):
\begin{equation} \label{ZEqnNum317320} D_{\mu \nu } \left(p\right)=\frac{1}{p^{2} +m^{2} } \left(g_{\mu \nu } +\frac{p_{\mu } p_{\nu } }{m^{2} } \right).  \end{equation}
Here $p_{\mu } $ are the components of the 4D energy-momentum vector, ${p^2} \equiv {g^{\mu \nu }}{p_\mu }{p_\nu } = \left( {{{\bar \omega }^2} + {{\vec k}^2}} \right)$, $\bar{\omega }=i\omega $ is the imaginary energy, and $m$ is the mass of the vector boson. In this case $m$is equal to the mass $m_{Z} \approx 92.5$~GeV of the neutral intermediate $Z^{0} $-boson. A relatively small decrease in the mass $m_{W} \approx 82.5$~GeV of charged $W^{\pm } $-bosons is due to their additional interaction with the electromagnetic field.

We note that the numerator of the Green function \eqref{ZEqnNum317320} determines the structure of the vacuum-vacuum matrix element for massive particles with spin $S=1$ and with vector (current) sources ${J^\mu }\left( p \right)$ \cite{Schwinger1970}:
\begin{multline} \label{ZEqnNum193060} \left\langle {{0_ + }} \right|{\left| {{0_ - }} \right\rangle ^J} = \\ \exp \left\{ {\frac{i}{2}\int {d{\omega _p}{J^\mu }{{\left( p \right)}^*}\left( {{g_{\mu \nu }} + \frac{{{p_\mu }{p_\nu }}}{{{m^2}}}} \right){J^\nu }\left( p \right)} } \right\},  \end{multline}
where $d\omega _{p} $ --- is an element of the phase volume in the 4D energy-momentum space.

For the transition from a microscopic field-theoretic description of the retarded weak interaction to the phenomenological Fermi theory, we use the convolution of the Green function \eqref{ZEqnNum317320} with a metric tensor $g^{\mu \nu } $ and, taking into account the normalization condition $g^{\mu \nu } g_{\mu \nu } =1$, obtain:
\begin{equation} \label{9} D_{V} \equiv g^{\mu \nu } D_{\mu \nu } ={1\mathord{\left/ {\vphantom {1 m^{2} }} \right. \kern-\nulldelimiterspace} m^{2} } .  \end{equation}

In accordance with the physical dimension of the Fermi constant $\left[ {{G_F}} \right] = \left[ {{{{e^2}} \mathord{\left/ {\vphantom {{{e^2}} {{m^2}}}} \right. \kern-\nulldelimiterspace} {{m^2}}}} \right]$, we introduce some effective ``charge'' of the weak interaction $e_{w} $ and represent this constant in the following form:
\begin{equation} \label{ZEqnNum301910} G_{F} =e_{w}^{2} \cdot D_{V} \equiv {e_{w}^{2} \mathord{\left/ {\vphantom {e_{w}^{2}  m_{Z}^{2} }} \right. \kern-\nulldelimiterspace} m_{Z}^{2} } .  \end{equation}

Taking into account the empirical values of the constant $G_{F} \approx 1.1664\cdot 10^{-5} $~GeV and the $Z^{0} $-boson mass, we obtain an estimate for the dimensionless weak-coupling constant $e_{w}^{2} \approx 10^{-1} $, which is determined by the relation $e_w^2 = {{{g^2}m_Z^2} \mathord{\left/ {\vphantom {{{g^2}m_Z^2} {4\sqrt 2 m_W^2}}} \right. \kern-\nulldelimiterspace} {4\sqrt 2 m_W^2}}$, where $g$ is one of the electroweak interaction constants having the electric charge dimension (the second constant is equal to $g'=g\cdot \tan\theta _{W} $, where $\theta _{W} $ is the Weinberg angle).

We note that in accordance with the asymptotic re\-normalization-group relation for coupling constants on a floating scale of energies (see \cite{Weinberg1996}), for the ratio of the constants of the weak $e_{w}^{2} $ and electromagnetic ${e^2} \approx {1 \mathord{\left/ {\vphantom {1 {137}}} \right. \kern-\nulldelimiterspace} {137}}$ interactions, the condition ${{e_w^2} \mathord{\left/ {\vphantom {{e_w^2} {{e^2}}}} \right. \kern-\nulldelimiterspace} {{e^2}}} \approx \ln \left( {{{{m_Z}} \mathord{\left/ {\vphantom {{{m_Z}} {{m_e}}}} \right. \kern-\nulldelimiterspace} {{m_e}}}} \right)$, where $m_{e} $ is the mass of the electron, is approximately satisfied.

As mentioned above, it was suggested in \cite{Pashitskii2017} that in spite of the anomalously small value of the ratio ${{{G_N}} \mathord{\left/ {\vphantom {{{G_N}} {{G_F}}}} \right. \kern-\nulldelimiterspace} {{G_F}}} \approx 6 \cdot {10^{ - 34}}$ there is a similarity between gravitational and weak interactions which lies in the assumption that just like the weak forces are caused by the exchange of virtual intermediate vector Bosons, the ``superweak'' gravitational forces are caused by the exchange of ``supermassive'' tensor bosons with a mass $M_{T} $ that is 17 orders of magnitude larger than the mass $m_{Z} $.

The vacuum-vacuum matrix element for massive tensor bosons with spin  $S=2$, which is determined by ten independent sources of a symmetric tensor field ${T^{\mu \nu }}\left( p \right) = {T^{\nu \mu }}\left( p \right)$, has the form \cite{Schwinger1970}:
\begin{multline} \label{11} \left\langle {{0_ + }} \right|{\left| {{0_ - }} \right\rangle ^T} = \\ \exp \left\{ {\frac{i}{2}\int {d{\omega _p}{{\bar T}^{\mu \nu }}{{\left( p \right)}^*}\left[ {{{\bar g}_{\mu \kappa }}\left( p \right) \cdot {{\bar g}_{\nu \lambda }}\left( p \right)} \right]{{\bar T}^{\kappa \lambda }}\left( p \right)} } \right\},  \end{multline}
where
\begin{equation} \label{12} {\bar T^{\mu \nu }}\left( p \right) = {T^{\mu \nu }}\left( p \right) - \frac{1}{3}{g^{\mu \nu }} \cdot \left[ {{{\bar g}_{\rho \sigma }}\left( p \right){T^{\rho \sigma }}\left( p \right)} \right];  \end{equation}
\begin{equation} \label{ZEqnNum347271} {\bar g_{\mu \nu }}\left( p \right) = {g_{\mu \nu }} + \frac{{{p_\mu }{p_\nu }}}{{{m^2}}} .  \end{equation}

It follows that, by analogy with \eqref{ZEqnNum317320} and \eqref{ZEqnNum193060}, the Green function of tensor bosons with mass $m=M_{T} $ has the following form:
\begin{equation} \label{ZEqnNum590997} D_{\mu \nu ,\kappa \lambda }^{T} \left(p\right)=\frac{\bar{g}_{\mu \nu } \left(p\right)\cdot \bar{g}_{\kappa \lambda } \left(p\right)}{p^{2} +M_{T}^{2} } .  \end{equation}

For a simplified phenomenological description of the retarded gravitational interaction due to the exchange of virtual massive tensor bosons, we consider the double convolution of the Green function \eqref{ZEqnNum590997} with metric tensors $g^{\mu \nu } $ and $g^{\kappa \lambda } $:
\begin{equation} \label{ZEqnNum847073} D_{T} \left(p\right)\equiv g^{\mu \nu } g^{\kappa \lambda } D_{\mu \nu ,\kappa \lambda }^{T} \left(p\right)=\frac{1}{M_{T}^{2} } \left(1+\frac{p^{2} }{M_{T}^{2} } \right).  \end{equation}

In the limit $p\to 0$, the gravitational constant, by analogy with \eqref{ZEqnNum301910}, can be represented as:
\begin{equation} \label{16} {G_N} = e_g^2 \cdot {D_T}\left( 0 \right) = {{e_g^2} \mathord{\left/ {\vphantom {{e_g^2} {M_T^2}}} \right. \kern-\nulldelimiterspace} {M_T^2}} ,  \end{equation}
where $e_{g} $ is the effective ``gravitational charge'' with the dimension of the electric charge.

Suppose that the coupling constants $e_{g}^{2} $ and $e_{w}^{2} $ given the condition $M_{T} \gg m_{Z} $, just as the constants $e_{w}^{2} $ and $e^{2} $ under the condition $m_{Z} \gg m_{e} $, are related to each other by the asymptotic logarithmic renormalization expression \cite{Weinberg1996}:
\begin{equation} \label{ZEqnNum158371} e_{g}^{2} =e_{w}^{2} \cdot \ln \left({M_{T} \mathord{\left/ {\vphantom {M_{T}  m_{Z} }} \right. \kern-\nulldelimiterspace} m_{Z} } \right).  \end{equation}

As a result, taking into account relations \eqref{ZEqnNum301910} and \eqref{ZEqnNum347271}, we obtain a transcendental equation for determining the mass ratio of tensor $M_{T} $ and intermediate $m_{Z} $ bosons (see \cite{Pashitskii2017}):
\begin{equation} \label{ZEqnNum982551} \left(\frac{M_{T} }{m_{Z} } \right)^{2} =\frac{G_{F} }{G_{N} } \cdot \ln \left(\frac{M_{T} }{m_{Z} } \right).  \end{equation}

Equation \eqref{ZEqnNum982551} has two roots, the smaller of which with the accuracy of about ${{{G_N}} \mathord{\left/
 {\vphantom {{{G_N}} {{G_F}}}} \right.
 \kern-\nulldelimiterspace} {{G_F}}} \approx 6 \cdot {10^{ - 34}}$ equals to unity and corresponds to the energies of the order of 100~GeV, which is typical for the standard model (SM) in the theory of elementary particles.

The second root of equation \eqref{ZEqnNum982551} equals to ${M_{T} \mathord{\left/ {\vphantom {M_{T}  m_{Z} }} \right. \kern-\nulldelimiterspace} m_{Z} } \approx 2.6\cdot 10^{17} $ and corresponds to the mass of tensor bosons $M_{T} \approx 2.4\cdot 10^{19} $~GeV, which is twice the Planck mass ${M_P} \equiv {1 \mathord{\left/
 {\vphantom {1 {\sqrt {{G_N}} }}} \right.
 \kern-\nulldelimiterspace} {\sqrt {{G_N}} }} \approx 1.2 \cdot {10^{19}}$~GeV.

Since, according to the results of \hyperref[Sec2]{Section~2}
, the mass of the tensor bosons is $M_{T} =\tilde{g}\varphi _{0} $ and the vacuum average of the scalar field equals to  ${\varphi _0} = {\mu  \mathord{\left/ {\vphantom {\mu  g}} \right. \kern-\nulldelimiterspace} g} = 2{M_P}\sqrt 2 $ provided the mass of the scalar bosons is $M_{B} =\mu \sqrt{2} \approx M_{P} $, we obtain an estimate $\tilde g \approx {1 \mathord{\left/ {\vphantom {1 {\sqrt 2 }}} \right. \kern-\nulldelimiterspace} {\sqrt 2 }}$ for the coupling constant of the scalar and gravitational fields in the Lagrangian \eqref{ZEqnNum805752}, if the vacuum energy density equals to ${\lambda _0} = {{{\mu ^2}\varphi _0^2} \mathord{\left/ {\vphantom {{{\mu ^2}\varphi _0^2} 4}} \right.
 \kern-\nulldelimiterspace} 4} \approx M_P^4$. In addition, when ${M_T} \approx 2M_{P}$, according to \eqref{ZEqnNum158371}, we obtain the following estimate for the dimensionless gravitational coupling constant: $e_{g}^{2} \approx 40e_{w}^{2} \approx 4$.

According to the relation \eqref{ZEqnNum847073}, the intensity of the gravitational interaction in the real time framework is determined by the function that depends on the energy $\omega $ and momentum $\vec{k}$:
\begin{equation} \label{ZEqnNum260909} {\tilde G_N}\left( {\omega ,\vec k} \right) = {G_N} \cdot \left( {1 - \frac{{{\omega ^2} - {{\vec k}^2}}}{{M_T^2}}} \right).  \end{equation}

In the static and long-wavelength limits, when $\omega \to 0$ and $\left|\vec{k}\right|\to 0$, the function \eqref{ZEqnNum260909} is reduced to the Newton's gravitational constant $G_{N} $. However, at small space-time scales, comparable to the Planck length and time scales ${l_P} = {t_P} = {1 \mathord{\left/ {\vphantom {1 {{M_P}}}} \right. \kern-\nulldelimiterspace} {{M_P}}}$, the value of $\tilde{G}_{N} $ may significantly differ from $G_{N} $.

At the same time, for any massless gauge or fermion fields with the excitation spectrum $\omega \left( {\vec k} \right) = \left| {\vec k} \right|$, the func\-tion \eqref{ZEqnNum260909} is identically equal to $G_{N} $ on all scales. On the other hand, for a massive tensor field with a spectrum of excitations
\begin{equation} \label{ZEqnNum953069} {\Omega _T}\left( {\vec k} \right) = \sqrt {M_T^2 + {{\left| {\vec k} \right|}^2}}  \end{equation}
the function \eqref{ZEqnNum260909} vanishes identically, i.e. the gravitational interaction is turned off. This means that for the propagation of virtual ``supermassive'' tensor bosons carrying ``superweak'' gravitational interaction the 4D space-time remains flat.

It should also be notes that the spectrum of quantum fluctuations of a vacuum filled with a fundamental scalar field with a scalar boson mass $_{} $ at an absolute zero of temperature is characterized by a $\delta $-shaped peak at a frequency
\begin{equation} \label{21} {\Omega _B}\left( {\vec k} \right) = \sqrt {M_B^2 + {{\left| {\vec k} \right|}^2}}.  \end{equation}

For such excitations with $M_{B} \ne M_{T} $ the expression \eqref{ZEqnNum260909} gives ${\tilde G_N} = {G_N} \cdot \left( {1 - {{M_B^2} \mathord{\left/ {\vphantom {{M_B^2} {M_T^2}}} \right. \kern-\nulldelimiterspace} {M_T^2}}} \right)$. In order to avoid the instability of the vacuum under the action of antigravity $\tilde{G}_{N} <0$, it is necessary for the condition $M_{B} \le M_{T} $ to be satisfied. In particular, for the values $M_{B} \approx M_{P} $ and $M_{T} \approx 2M_{P} $ given above, we obtain an estimate ${\tilde G_N} \approx {{3{G_N}} \mathord{\left/ {\vphantom {{3{G_N}} 4}} \right. \kern-\nulldelimiterspace} 4}$.

At the same time, it should be emphasized that under the condition $M_{B} =M_{T} $ the function \eqref{ZEqnNum260909} is identically equal to zero on all scales, so in this case the gravitational interaction completely disappears and the 4D space-time becomes flat.

\section{Conclusion}

Thus, it follows from the foregoing that it is quite feasible to construct a renormalizable quantum field theory of gravity in the initially flat 4D space-time without involving additional dimensions. It becomes clear that the tensor equations of general relativity are a phenomenological way of describing gravity as the curvature of the 4D space-time, taking into account the principle of equivalence of inertial and gravitational masses. These equations are valid only on sufficiently large space-time scales much exceeding the Planck scales $l_{P} $ and $t_{P} $. At the same time, Newton's dimensional gravitational constant $G_{N} $ is an approximate phenomenological constant in the domain of energy and momentums that are small in comparison with the Planck values of energy and momentum $M_{P} $, analogously to the phenomenological Fermi constant $G_{F} $, which is constant only in the region of low energies and small momenta much smaller than the value characteristic of electroweak interaction, which is determined by the mass of the Higgs boson $m_{H} \approx 100$~GeV.

\section*{Acknowledgments}
We would like to express our gratitude to V.P.~Gusynin and Yu.A.~Sitenko for useful discussions of some of the issues considered in this paper.

\section*{References}

\bibliography{GravConstElsevier}

\end{document}